**Bayesian Markov Chain Monte Carlo-Based Simultaneous Speed Prediction Model for Heterogeneous Traffic on Urban Primary Roads**

**S.M Towhidul Alam**
Graduate Research Assistant
Department of Civil Engineering
Bangladesh University of Engineering and Technology (BUET), Dhaka, Bangladesh, 1000
Tel: +880-2-55167100; Email: towhid6320@gmail.com

**Md. Muhtashim Shahrier**
Lecturer
Department of Civil Engineering
Bangladesh University of Engineering and Technology (BUET), Dhaka, Bangladesh, 1000
Cell: +880-1729249589; Email: shahriermuhtashim@gmail.com

**Nazmul Haque**
Assistant Professor
Accident Research Institute (ARI),
Bangladesh University of Engineering and Technology (BUET), Dhaka, Bangladesh, 1000
Tel: +8801673174808; Fax: +880-2-58610081; Email: nhaque@ari.buet.ac.bd

**Md Asif Raihan***
Professor
Accident Research Institute (ARI)
Bangladesh University of Engineering and Technology (BUET), Dhaka, Bangladesh, 1000
Cell: + 8801911142802; Email: raihan@ari.buet.ac.bd

**Md. Hadiuzzaman**
Professor
Department of Civil Engineering
Bangladesh University of Engineering and Technology (BUET), Dhaka, Bangladesh, 1000
Tel: 88-02-55167100 Ext. 7225, Fax: 880-2-58613046; Email: mhadiuzzaman@ce.buet.ac.bd

*Corresponding Author

**ABSTRACT**

**Objectives:** The study designs a simultaneous speed prediction model based on Bayesian Markov Chain Monte Carlo (MCMC) methods for heterogeneous, non-lane-based traffic on urban primary roads, with the aim of determining the effects of vehicle-specific traffic densities on the operating speeds of the six vehicle types.

**Methods:** The traffic data were obtained from videos recorded by unmanned aerial vehicles on five main urban roads in Dhaka, Bangladesh, and were then processed by a deep learning approach in order to derive information on vehicle trajectories, traffic volumes, densities, and space-mean speeds. Six simultaneous equations for speed prediction were developed using a Bayesian MCMC framework, covering standard cars, utility vehicles, heavy vehicles, three-wheelers, two-wheelers, and non-motorized vehicles. The performance of the models was assessed by means of goodness-of-fit measures.

**Findings:** The coefficients of determination for the models varied from 0.922 to 0.956 and the root mean square errors were between 0.50 and 1.56 km/h for the six types of vehicle. Operating speeds decreased steadily as traffic volume increased, although traffic composition had a significant effect on speed when the demand was moderate or high. Operating speeds increased and congestion was delayed when non-motorized vehicles were excluded. The transferability study showed that the models had a reliable ability to make predictions on separate road sections.

**Novelty:** This study introduces the first Bayesian MCMC-based method for simultaneously predicting macroscopic speeds in heterogeneous, non-lane-based urban traffic, using data obtained from unmanned aerial vehicles; the approach involves estimating the posterior parameter distributions alongside a model of the interactions between six vehicle categories.

**Practical Applications:** The model suggested serves as a decision-support tool for assessing traffic performance when traffic demand and vehicle composition vary. It is capable of aiding in traffic management, the analysis of roadway operations, traffic simulation, and infrastructure planning in situations involving heterogeneous traffic.

## INTRODUCTION

Urban mobility is a key determinant of economic productivity, accessibility, and quality of life. Rapid urbanization has substantially increased travel demand, placing considerable pressure on urban transportation infrastructure and resulting in recurrent congestion and declining operational performance. Effective traffic management therefore depends on reliable traffic performance models capable of accurately representing traffic conditions under varying roadway and traffic environments (Williams 1997).

Operating speed is one of the most important indicators of traffic performance because it directly reflects roadway operating conditions while influencing capacity, travel time, fuel consumption, and traffic safety (Dhamaniya and Chandra 2013). As one of the three fundamental variables of traffic flow theory, together with flow and density, operating speed is widely used in traffic simulation, roadway design, signal optimization, Intelligent Transportation Systems (ITS), and Highway Capacity Manual (HCM)-based level-of-service evaluation.

Reliable speed prediction remains particularly challenging in low- and middle-income countries (LMICs), where traffic streams are highly heterogeneous and lane discipline is weak. Unlike the homogeneous traffic assumed in conventional traffic flow theory, urban roads accommodate diverse vehicle classes—including passenger cars, utility vehicles, heavy vehicles, motorcycles, three-wheelers, and non-motorized vehicles—that continuously compete for the same roadway space. Frequent lateral movements, overtaking maneuvers, and substantial speed differentials create complex vehicle interactions that cannot be adequately represented by conventional macroscopic speed–density relationships. Dhaka, Bangladesh, exemplifies such traffic conditions, where heterogeneous traffic and weak lane discipline substantially influence operating speeds on urban primary roads.

Various macroscopic speed prediction models have been suggested in previous studies. Greenshields et al. (1935) set up the classical speed–density relationship which serves as the foundation for traffic flow modelling. Dhamaniya and Chandra (2013) showed that speed prediction equations which are solved simultaneously lead to a better representation of the interactions between different vehicle types in mixed traffic. Munawar (2011) pointed out the effect of the heterogeneous composition of vehicles on urban traffic performance, Tan et al. (2021) used Maximum Likelihood Estimation to enhance parameter estimation in mixed traffic, and Ahmed et al. (2021) demonstrated, by means of unmanned aerial vehicle (UAV)-derived trajectory data, that traffic heterogeneity greatly changes the conventional speed–flow relationships.

Even though there have been such advances, current simultaneous speed prediction models mainly focus on point estimation and give only a limited measure of the uncertainty connected with interactions among different types of traffic. Bayesian inference provides a rigorous probabilistic approach since it estimates the complete posterior distributions of the model parameters and thus explicitly accounts for uncertainty while taking into consideration the stochastic character of heterogeneous traffic. Yet, Bayesian simultaneous macroscopic speed prediction models for multiple interacting vehicle categories based on high-resolution UAV-derived traffic data have not been created for urban primary roads which operate under weak lane discipline.

Therefore, the present study is intended to (i) create Bayesian simultaneous macroscopic speed prediction models for six major types of vehicles by making use of traffic data obtained from UAVs, (ii) determine the model parameters via a Bayesian Markov Chain Monte Carlo (MCMC) framework, (iii) assess the models' performance and transferability by means of independent road data, and (iv) examine the impact of traffic volume and traffic composition on operating speeds by means of a sensitivity analysis. Traffic data gathered from five major urban roads in Dhaka were processed using the DEEGITS deep learning framework (Islam et al. 2024). The framework proposed explicitly takes into account parameter uncertainty while at the same time modelling the interactions between the six vehicle categories, thus providing a transferable method for predicting speeds on heterogeneous, non-lane-based urban road networks.

## RELEVANT WORKS

Macroscopic speed prediction has been studied in great detail since vehicle speed is a major measure of traffic performance and is essential for traffic management, capacity analysis, and transportation planning. The first speed–density relationship was put forward by Greenshields et al. in 1935, who proposed a linear speed–density model which then became the foundation of the classical speed–flow relationship. Even though the model is simple and computationally efficient, it assumes that the traffic is homogeneous and that there is strict lane discipline, which limits its use in heterogeneous urban traffic conditions (Greenshields et al. 1935).

Later on, research has been directed towards enhancing the prediction of operating speed by taking into account roadway geometry, traffic flow, and vehicle characteristics. Donnell et al. (2001) created operating speed models for trucks on two-lane rural highways, and Himes and Donnell (2010) used a simultaneous equation framework in order to show the endogenous relationship between mean operating speed and speed variability on multilane highways. Transportation Research Circular E-C151 also stressed that conventional linear regression models usually do not succeed in depicting the complex interactions among traffic variables due to their assumptions about data independence and linearity, and suggested that more sophisticated modelling methods should be used for the analysis of operating speed.

Most of the research into heterogeneous traffic has been carried out in developing countries, where mixed traffic and a lack of strict lane discipline are common. Munawar (2011) showed that operating speed decreases slowly at low traffic volumes but drops sharply as traffic nears the capacity of the road on urban roads in Indonesia. Likewise, Dey et al. (2006) found that heterogeneous traffic results in multimodal speed distributions due to the presence of both slow and fast-moving vehicles. Nagaraj et al. (1990) linked operating speed to the lateral clearance requirements, and Sarna et al. (1989) noted that a higher proportion of auto-rickshaws leads to a reduction in stream speed. Moreover, Maini and Khan (2000) pointed out that in congested conditions, the interactions between different types of vehicles cause the traffic to act as one continuous group moving at relatively uniform speeds.

In the case of studies that focused on mixed traffic, Dhamaniya and Chandra (2013) suggested a set of simultaneous speed prediction equations according to which the speed of each vehicle type is given as a function of the densities of all the vehicle classes. Their research showed that simultaneous modelling offers a more realistic depiction of the interactions among heterogeneous traffic types than does modelling based on aggregate speeds and thus provided a valuable framework for predicting macroscopic speeds on a vehicle-by-vehicle basis.

Progress in sensing technologies and statistical modelling has also led to further improvements in the analysis of heterogeneous traffic. Ahmed et al. (2021) used trajectory data obtained from UAVs to show that weak lane discipline and vehicle heterogeneity greatly change the fundamental speed–density relationship as compared to traffic analysis on a lane-by-lane basis. Tan et al. (2021) applied Maximum Likelihood Estimation in order to estimate speed models for mixed traffic in Hanoi, while Najjar et al. (2000) used artificial neural networks to predict operating speeds with a high degree of accuracy. More recently, machine learning and deep learning techniques have been taken up more and more for use in traffic prediction. Studies carried out by Liao et al. (2018), Monteil et al. (2021), Du et al. (2021), and Alomari et al. (2022) proved that advanced learning algorithms are able to increase prediction accuracy by taking into account temporal, spatial and contextual information. Even though there have been some advances, a relatively small number of studies have developed macroscopic speed prediction models that are specifically designed for highly heterogeneous, non-lane-based urban traffic. The existing simultaneous speed models have so far been mainly created for urban arterials in India, whereas comparable models for Dhaka's urban primary roads are still not available. Moreover, earlier research has seldom used a Bayesian inference framework to estimate simultaneous vehicle-class-specific speed models and has not explicitly quantified parameter uncertainty. Because of these limitations, there is a need to develop a simultaneous speed prediction model based on Bayesian Markov Chain Monte Carlo (MCMC) that is capable of capturing the complex interactions among different vehicle categories in heterogeneous urban traffic conditions.

## METHODOLOGY

### Model Formulation

The speed at which vehicles operate on a road is affected by the road's geometry, the environmental conditions, and the current traffic characteristics. As the geometric design and the surrounding environmental conditions do not change significantly over the period of a field observation, any variations in operating speed can mainly be explained by changes in traffic conditions. Of these factors, traffic density, traffic volume, and the make-up of the vehicles are the main factors that determine the speed of the traffic stream.

With an increase in the amount of traffic, there are more interactions between vehicles, which in turn leads to a higher traffic density and greater variations in both the speed of the traffic stream and the speeds of individual vehicles. In heterogeneous traffic, these interactions are made more complicated by the fact that a number of different types of vehicles, each with its own physical dimensions and operating characteristics, are present together. Since the make-up of the traffic changes, the intensity of the interactions also changes because different kinds of vehicles take up different amounts of road space and have different abilities to accelerate, decelerate and change direction. As a result, the effect of traffic density on speed depends not just on the total number of vehicles but also on the composition of the traffic stream.

The relationship between speed and density has traditionally been represented by the classical Greenshields model, expressed as in **Equation 1**

$$V = V_f\left(1 - \frac{K}{K_j}\right) \tag{1}$$

Where The average traffic speed is denoted by, the free-flow speed by $V_f$, the traffic density by, and the jam density by $K_j$. For model estimation, **Equation 1** can be written in the equivalent linear form.

$$V = a - bK \tag{2}$$

where $a$and $b$ are the regression parameters, with $a$ representing the free-flow speed and $b$ describing the rate at which the speed decreases as the traffic density increases.

The Greenshields formulation is widely used in the case of homogeneous traffic, but its direct use in relation to heterogeneous, non-lane-based traffic is restricted since the total traffic density includes contributions from various types of vehicles. Because each vehicle category differs in size, dynamic performance, and demand for roadway space, it has a different effect on the speeds of the vehicles nearby. As a result, it is not possible to represent the traffic stream by means of a single average density if one is to adequately take into account these heterogeneous interactions.

In order to take these effects into account, the total traffic density is broken down into the densities of the various types of vehicles. The speed of each vehicle class is then treated as a function of the densities of all the vehicle categories present in the traffic stream. It is thus possible to express the generalized speed-density relationship as in **Equation 3**.

$$V_i = a_i - \sum_{j=1}^{N} b_{ij} K_j \tag{3}$$

Where $V_i$ is the operating speed of vehicle category $i$, $K_j$ denotes the density of vehicle category $j$, and $N$represents the total number of vehicle categories considered in the analysis. The parameter $a_i$corresponds to the free-flow speed of vehicle category $i$, while $b_{ij}$ quantifies the influence of the density of vehicle category $j$ on the speed of vehicle category $i$.

This formulation allows for the creation of a set of simultaneous equations for predicting speed, with each vehicle category having its own equation. Because the operating speed of each vehicle class is affected by the densities of all vehicle categories, the resulting model reflects the mutual interactions between the different traffic components and offers a more realistic depiction of mixed traffic conditions than do conventional aggregate speed-density models.

**Markov Chain Monte Carlo-Based Parameter Estimation**

The speed prediction models were developed using a Bayesian inference approach with the aid of Markov Chain Monte Carlo (MCMC) simulation (Brooks 1998). In the Bayesian method, the model parameters are regarded as random variables described by probability distributions, not as fixed but unknown values. As a result, parameter estimation is carried out on the basis of posterior probability distributions, which enables the uncertainty linked to varying traffic conditions to be clearly quantified. This type of probabilistic approach is especially appropriate for non-lane-based mixed traffic situations since the complex interactions among different types of vehicles lead to a great deal of variation in operating speeds. For each vehicle category, the simultaneous speed prediction model is expressed as in **Equation 4**.

$$V_i = a_i - \sum_{j=1}^{N} b_{ij} K_j + \varepsilon_i \tag{4}$$

where $V_i$ denotes the operating speed of vehicle category $i$, $K_j$ represents the density of vehicle category $j$, $a_i$ is the free-flow speed (intercept), $b_{ij}$ denotes the effect of the density of vehicle category $j$ on the speed of vehicle category $i$, and $\varepsilon_i$ is the random error term. The residual errors are assumed to follow an independent normal distribution,

$$\varepsilon_i \sim N(0, \sigma_i^2) \tag{5}$$

where $\sigma_i^2$ is the residual variance associated with the corresponding speed equation. Under the normality assumption, the likelihood function for the observed speed data is defined as

$$P(V \mid \theta) = \prod_{n=1}^{m} \frac{1}{\sqrt{2\pi\sigma^2}} \exp\left[ -\frac{\left(V_n - \hat{V}_n\right)^2}{2\sigma^2} \right] \tag{6}$$

where $V_n$ and $\hat{V}_n$ denote the observed and predicted speeds for observation $n$, respectively, $m$ is the total number of observations, and $\theta$ represents the complete set of unknown model parameters.

Bayesian inference combines the likelihood with prior information through Bayes' theorem to obtain the posterior distribution of the parameters,

$$(\theta \mid V) = \frac{P(V \mid \theta) P(\theta)}{P(V)} \tag{7}$$

where $P(\theta)$ denotes the prior distribution, $P(V \mid \theta)$ is the likelihood function, and $P(V)$ is the marginal likelihood acting as a normalizing constant. Weakly informative normal priors were assigned to the regression coefficients to provide numerical stability while allowing the observed traffic data to dominate the posterior inference. A positive prior distribution was introduced in order to make sure that the variance estimates would be physically meaningful.

Since there is no closed-form analytical solution for the posterior distribution in the system of equations, a numerical approximation was obtained by means of a Markov Chain Monte Carlo (MCMC) simulation. This method involves creating a Markov chain whose stationary distribution approaches the desired posterior distribution; after undergoing an initial burn-in period, samples from the posterior distribution were gathered and employed to estimate the posterior means, the posterior standard deviations, and the 95% credible intervals for all of the model parameters.

The statistical description of the simultaneous speed prediction models obtained through the use of MCMC includes an explicit representation of parameter uncertainty. The posterior estimates thus obtained were then employed to assess the significance of the explanatory variables, to quantify the uncertainty associated with speed prediction, and to carry out sensitivity and transferability analyses for different heterogeneous traffic compositions. The Bayesian MCMC framework therefore offers a solid foundation for estimating simultaneous speed prediction models in complex, non-lane-based urban traffic environments.

**Solution of the Simultaneous Speed Prediction System**

After having carried out the parameter estimation, the simultaneous speed prediction equations were used to estimate the operating speeds of all vehicle categories for the specified traffic demand and composition scenarios. Unlike conventional regression models, the proposed formulation is inherently coupled because traffic density depends on the unknown operating speed of each vehicle category. Consequently, the estimated regression equations form a nonlinear system that must be solved simultaneously.

For a given traffic condition, the flow of each vehicle category, $q_j$, was determined from the total traffic demand and the corresponding traffic composition. The density of each vehicle category was then computed using the fundamental traffic flow relationship

$$K_j = \frac{q_j}{V_j} \tag{8}$$

where $K_j$, $q_j$, and $V_j$ denote the density, flow, and operating speed of vehicle category $j$, respectively. Substituting **Equation 8** into the simultaneous speed prediction model yields

$$V_i = a_i - \sum_{j=1}^{N} b_{ij} \left( \frac{q_j}{V_j} \right), i = 1, \dots, N \tag{9}$$

where the operating speed of each vehicle category depends on the unknown speeds of all vehicle classes. Since $V_j$ appears in the denominator, the resulting system is nonlinear and does not admit a closed-form analytical solution. The unknown speeds were therefore obtained by solving the nonlinear simultaneous system through numerical optimization. Let $\hat{V}_i$ denote the speed predicted by the simultaneous equation and $V_i$ the corresponding optimization variable. The solution was obtained by minimizing the root mean square error (RMSE) of the residuals across all simultaneous equations,

$$\min_{\mathbf{V}} \sqrt{\frac{1}{N}\sum_{i=1}^{N} E_i^2} \tag{10}$$

Where

$$E_i = V_i - \left(a_i - \sum\nolimits_{j=1}^{N} b_{ij}\frac{q_j}{V_j}\right) \tag{11}$$

is the residual associated with the $i^{\text{th}}$ speed prediction equation.

The optimization problem was solved using a constrained nonlinear optimization algorithm, where physically realistic lower and upper bounds were imposed on the operating speeds of all vehicle categories. To reduce the possibility of convergence to local optima resulting from the highly nonlinear objective function, a MultiStart optimization strategy was adopted. Multiple optimization runs were initiated from different starting points, and the solution corresponding to the minimum objective value was selected as the final estimate. This procedure ensures a stable and consistent solution of the simultaneous speed prediction system across different traffic conditions.

After having determined the optimal speeds, the relevant traffic densities were recalculated with the help of **Equation 8**. The proposed framework was able to obtain equilibrium operating speeds for all vehicle types under various traffic scenarios by altering the total traffic demand while keeping the specified traffic composition ratios unchanged. The speeds that were predicted in this way were then used to examine the impact of traffic demand and vehicle composition on the operating performance of heterogeneous urban traffic streams.

## DATA COLLECTION AND PROCESSING

The traffic data were obtained by using a DJI Air 3 drone fitted with a dual-camera system that could record HDR video at 4K resolution and 60 frames per second. The UAV was flown at a constant altitude so as to keep a steady ground sampling distance and thus achieve uniform spatial resolution over all the areas studied. Various important factors were taken into account in order to guarantee the quality and reliability of the data collected by means of the UAV (Khan et al. 2017). Collection of the data took place only on days that were clear and had light winds in order to avoid image instability. By following nadir flight paths perspective distortion was reduced and the use of several batteries ensured that the areas could be covered continuously within the available flight time. Compliance with the relevant urban aviation regulations in Bangladesh was maintained for all UAV operations.

Five urban primary road segments (1 to 5) in Dhaka city were selected for data collection: Mirpur Road, Matikata Road, Airport Road, Toyenbee Circular Road, and Darus Salam Road. Each segment was selected based on uniform traffic operating conditions, absence of intersections, driveways, or access points within the study boundary, and representation of varied but typical heterogeneous traffic conditions found on Dhaka's primary road network. A 400-meter section was identified along each road, with a 60-meter longitudinal trap established at the mid-section for flow and speed measurement. The geometry of the selected road segments is summarized in **Table 1**.

**TABLE 1 Road Geometry of Selected Primary Road Segments**

| Road | Carriageway Width (m) | No. of Lanes | Lane Width (m) | Footpath | Footpath Width (m) |
|---|---|---|---|---|---|
| Mirpur Road | 14.68 | 4 | 3.67 | Yes | 3.0 |
| Matikata Road | 14.7 | 4 | 3.675 | Yes | 3.5 |
| Airport Road | 14.7 | 4 | 3.675 | Yes | 2.42 |

| Road | Carriageway Width (m) | No. of Lanes | Lane Width (m) | Footpath | Footpath Width (m) |
|---|---|---|---|---|---|
| Toyenbee Circular Road | 14.7 | 4 | 3.675 | Yes | 2.45 |
| Darus Salam Road | 14.7 | 4 | 3.675 | Yes | 3.5 |

Data collection was conducted on typical weekdays on from 8:00 a.m. to 8:00 p.m., covering a full twelve-hour period to capture the complete daily traffic demand cycle including morning peak, midday off-peak, and evening peak conditions. For each study segment, the UAV video footage recorded all vehicles passing through the 60-meter longitudinal trap. For each observation interval the vehicle counts, classifications and exact positions were taken from the video. The vehicle categories used those of the six-group classification system—Standard Cars (SC), Utility Vehicles (UV), Heavy Vehicles (HV), Three-Wheelers (3W), Two-Wheelers (2W), and Non-Motorized Vehicles (NMV).

The UAV video footage was analysed using the DEEGITS framework (Islam et al. 2024). The YOLOv8 algorithm detected and classified vehicles frame by frame, while DeepSORT was used to track the individual vehicle paths from one frame to the next. Kalman filtering was employed to smooth the estimated trajectories and to deal with short periods of occlusion resulting from vehicles overlapping — a common situation in dense traffic where there are no lanes. The space-mean speed for each vehicle category was calculated based on the times it took individual vehicles to travel the 60-metre longitudinal trap, in accordance with the theoretical definition of space-mean speed as the harmonic mean of the speeds of individual vehicles over a given spatial distance (Turner 1998). Any observations with incomplete trajectory data because of prolonged occlusion or vehicles entering or leaving the frame during the trap period were excluded from the analysis. Missing values that occurred during low-volume periods for certain vehicle categories were dealt with by interpolation in order to ensure the continuity of the dataset.

**Table 2** gives the speed statistics for vehicles in the processed traffic data, with the figures compiled for all five road sections and grouped by type of vehicle. The data illustrates the typical speed hierarchy found in Dhaka's mixed traffic — motorized vehicles travelling at greatly higher speeds than non-motorized ones, together with a large amount of variation within each vehicle category which is due to the traffic flow not being based on lanes and not being regulated. **Table 3** displays the 12-hour volumes of vehicles broken down by vehicle category.

**TABLE 2 Speed Statistics by Vehicle Category**

| Vehicle Category | Mean Speed (km/h) | Speed Range (km/h) |
|---|---|---|
| Standard Car (SC) | 24.5 | 12.9 – 51.0 |
| Utility Vehicle (UV) | 7.5 | 1.1 – 19.3 |
| Heavy Vehicle (HV) | 12.2 | 0.6 – 30.9 |
| Three-Wheeler (3W) | 19.3 | 11.9 – 30.8 |
| Two-Wheeler (2W) | 18.4 | 2.4 – 30.8 |
| Non-Motorized Vehicle (NMV) | 6.6 | 0.0 – 13.0 |

**TABLE 3 Vehicle Category Wise 12-hour Volume and Categorical Composition**

| Location | | SC | UV | HV | 3W | 2W | NMV | Total |
|---|---|---|---|---|---|---|---|---|
| Mirpur Road | # | 7912 | 7558 | 4315 | 6521 | 5288 | 2875 | 34469 |
| | % | 22.95 | 21.92 | 12.52 | 18.92 | 15.34 | 8.34 | |
| Matikata Road | # | 13599 | 8321 | 2415 | 4409 | 4687 | 2305 | 35736 |
| | % | 38.05 | 23.28 | 6.76 | 12.34 | 13.12 | 6.45 | |
| Airport Road | # | 16725 | 17456 | 7267 | 6704 | 6380 | 1158 | 55690 |
| | % | 30.03 | 31.34 | 13.05 | 12.04 | 11.46 | 2.08 | |
| Toyenbee Road | # | 4339 | 2643 | 539 | 3725 | 2761 | 6648 | 20655 |
| | % | 21.01 | 12.80 | 2.61 | 18.03 | 13.37 | 32.19 | |
| Darus Salam Road | # | 5368 | 4588 | 2249 | 2286 | 5451 | 11205 | 31147 |
| | % | 17.23 | 14.73 | 7.22 | 7.34 | 17.50 | 35.97 | |

## MODEL DEVELOPMENT

### Bayesian Parameter Estimates

The speed prediction models for the six types of vehicle were obtained by using the Bayesian MCMC framework as described in the 'METHODOLOGY' section. The posterior summaries, which include the posterior means, the standard deviations, the 95% credible intervals and the posterior probabilities, are given in **Table 4**. In all six models the MCMC chains showed stable convergence and the estimated coefficients had narrow credible intervals, showing that the parameter estimation was reliable. The values of the intercepts correspond to the free-flow speeds of the respective vehicle categories, while the regression coefficients measure the effect of the vehicle-category-specific densities on running speed.

The models developed showed a strong ability to make predictions for all types of vehicle. The coefficient of determination ($R^2$) was between 0.922 and 0.956, and the root mean square error (RMSE) was in the range of 0.50 to 1.56 m/s. The two-wheeler (2W) model had the best goodness-of-fit (with an $R^2$of 0.956), then came the standard car (SC) and the three-wheeler (3W) models (both having an $R^2$ of 0.947). The heavy vehicle (HV) model had the lowest value of the coefficient of determination ($R^2$=0.922), although this was still acceptable. Of all the vehicle categories, the non-motorized vehicle (NMV) model had the smallest prediction error (RMSE = 0.50 m/s), while the standard car (SC) model had the highest RMSE (1.56 m/s). These statistics show that the Bayesian simultaneous modeling framework proposed is able to accurately capture the speed characteristics of heterogeneous traffic and gives a reliable foundation for further analyses of the effects of traffic volume and traffic composition.

**TABLE 4 Consolidated Bayesian Posterior Statistics for All Vehicle Types**

| Model term | Mean (SD) | 95% CI | P(>0) | Model term | Mean (SD) | 95% CI | P(>0) |
|---|---|---|---|---|---|---|---|
| **$V_{SC}$: a, Intercept** | 13.863 (0.033) | [13.798, 13.927] | 1.000 | **$V_{3W}$: a, Intercept** | 10.351 (0.025) | [10.301, 10.401] | 1.000 |
| $b_1$ (SC) | -55.750 (1.023) | [-57.755, -53.745] | 0.000 | $b_1$ (SC) | -32.638 (0.765) | [-34.139, -31.138] | 0.000 |
| $b_2$ (UV) | -12.264 (0.480) | [-13.205, -11.324] | 0.000 | $b_2$ (UV) | -8.745 (0.395) | [-9.520, -7.970] | 0.000 |
| $b_3$ (HV) | -3.107 (0.210) | [-3.519, -2.695] | 0.000 | $b_3$ (HV) | -2.381 (0.163) | [-2.700, -2.061] | 0.000 |
| $b_4$ (3W) | -13.426 (0.983) | [-15.354, -11.499] | 0.000 | $b_4$ (3W) | -22.819 (0.751) | [-24.290, -21.347] | 0.000 |
| $b_5$ (2W) | -1.953 (0.770) | [-3.461, -0.445] | 0.006 | $b_5$ (2W) | -7.665 (0.577) | [-8.796, -6.533] | 0.000 |
| $b_6$ (NMV) | -2.743 (0.257) | [-3.247, -2.240] | 0.000 | $b_6$ (NMV) | -0.973 (0.109) | [-1.187, -0.759] | 0.000 |
| $\sigma^2$ | 2.446 (0.051) | [2.348, 2.549] | 1.000 | $\sigma^2$ | 1.462 (0.031) | [1.403, 1.523] | 1.000 |
| **$R^2$** | 0.947 | | | **$R^2$** | 0.947 | | |

| Model term | Mean (SD) | 95% CI | P(>0) | Model term | Mean (SD) | 95% CI | P(>0) |
|---|---|---|---|---|---|---|---|
| **RMSE** | 1.56 | | | **RMSE** | 1.17 | | |
| **$V_{UV}$: a, Intercept** | 5.857 (0.018) | [5.822, 5.892] | 1.000 | **$V_{2W}$: a, Intercept** | 9.899 (0.023) | [9.853, 9.945] | 1.000 |
| $b_1$ (SC) | -13.041 (0.539) | [-14.097, -11.985] | 0.000 | $b_1$ (SC) | -28.033 (0.707) | [-29.418, -26.648] | 0.000 |
| $b_2$ (UV) | -9.189 (0.278) | [-9.734, -8.643] | 0.000 | $b_2$ (UV) | -9.736 (0.365) | [-10.451, -9.021] | 0.000 |
| $b_3$ (HV) | -2.469 (0.115) | [-2.694, -2.244] | 0.000 | $b_3$ (HV) | -2.098 (0.150) | [-2.393, -1.804] | 0.000 |
| $b_4$ (3W) | -8.232 (0.528) | [-9.268, -7.196] | 0.000 | $b_4$ (3W) | -11.732 (0.693) | [-13.090, -10.374] | 0.000 |
| $b_5$ (2W) | -3.982 (0.406) | [-4.779, -3.185] | 0.000 | $b_5$ (2W) | -12.708 (0.533) | [-13.752, -11.663] | 0.000 |
| $b_6$ (NMV) | -0.487 (0.077) | [-0.637, -0.336] | 0.000 | $b_6$ (NMV) | -1.670 (0.101) | [-1.868, -1.472] | 0.000 |
| $\sigma^2$ | 0.724 (0.015) | [0.695, 0.755] | 1.000 | $\sigma^2$ | 1.246 (0.026) | [1.196, 1.298] | 1.000 |
| **$R^2$** | 0.924 | | | **$R^2$** | 0.956 | | |
| **RMSE** | 0.794 | | | **RMSE** | 1.016 | | |
| **$V_{HV}$: a, Intercept** | 10.276 (0.030) | [10.217, 10.336] | 1.000 | **$V_{NMV}$: a, Intercept** | 10.578 (0.037) | [10.506, 10.649] | 1.000 |
| $b_1$ (SC) | -35.157 (0.915) | [-36.950, -33.365] | 0.000 | $b_1$ (SC) | -42.091 (1.105) | [-44.256, -39.926] | 0.000 |
| $b_2$ (UV) | -9.219 (0.472) | [-10.145, -8.293] | 0.000 | $b_2$ (UV) | -9.200 (0.570) | [-10.318, -8.082] | 0.000 |
| $b_3$ (HV) | -7.159 (0.195) | [-7.540, -6.777] | 0.000 | $b_3$ (HV) | -0.001 (0.235) | [-0.462, 0.459] | 0.498 |
| $b_4$ (3W) | -12.391 (0.897) | [-14.149, -10.632] | 0.000 | $b_4$ (3W) | -10.562 (1.083) | [-12.685, -8.439] | 0.000 |
| $b_5$ (2W) | -6.196 (0.690) | [-7.548, -4.844] | 0.000 | $b_5$ (2W) | -9.138 (0.833) | [-10.771, -7.505] | 0.000 |
| $b_6$ (NMV) | -0.719 (0.131) | [-0.975, -0.463] | 0.000 | $b_6$ (NMV) | -1.692 (0.158) | [-2.001, -1.383] | 0.000 |
| $\sigma^2$ | 2.087 (0.044) | [2.004, 2.175] | 1.000 | $\sigma^2$ | 3.044 (0.064) | [2.922, 3.171] | 1.000 |
| **$R^2$** | 0.922 | | | **$R^2$** | 0.928 | | |
| **RMSE** | 1.40 | | | **RMSE** | 0.50 | | |

**Interpretation of Bayesian Posterior Coefficients**

The posterior estimates given in Table 6 have a clear physical interpretation in relation to heterogeneous, non-lane-based traffic. The intercept term for each vehicle type stands for the theoretical free-flow speed of that type of vehicle when all the density ratios tend to zero — in other words, when there is negligible traffic. Each β coefficient indicates the amount by which the speed of that vehicle type is reduced for each unit rise in the density ratio of the relevant vehicle category. The sign, size, and credible interval of each coefficient together explain the character and degree of the interactions between vehicles in the traffic stream.

**Standard Cars ($V_{SC}$).** The fact that the intercept is 13.863 m/s, or about 49.9 km/h, shows that this is the highest possible free-flow speed among the types of vehicles considered, something which is in agreement with the manoeuvrability and operating behaviour of standard cars on main roads. The self-density coefficient, $\beta_1$ = -55.750, is the most negative of all the coefficients in the system and means that as the density of SC increases there is an extremely strong decrease in SC speed; this is probably due to the large number of cars and the dense longitudinal and lateral interactions they generate when lane discipline is weak. The density of three-wheelers has the second greatest effect on SC speed, with $\beta_4$ = -13.426, implying that slower three-wheelers taking up the same amount of road space greatly disrupt the movement of cars. The effect of two-wheelers is smaller, $\beta_5$ = -1.953, and is associated with a comparatively wider range of credible values lying between [-3.461, -0.445], showing a higher level of uncertainty. It is possible that motorcycles move through the gaps that are available rather than continuously blocking the paths of cars. The effect of NMV is also small, $\beta_6$ = -2.743, although its narrow interval of [-3.247, -2.240] shows that it has a consistent slowing effect. The residual variance, $\sigma^2$ = 2.446, shows a moderate amount of unexplained variability.

**Utility Vehicles ($V_{UV}$).** The value of 5.857 m/s, or about 21.1 km/h, is considerably less than that observed for SCs, this being due to the larger size and slower performance of utility vehicles in heavy traffic. The self-density coefficient, $\beta_2$ = -9.189, shows a moderate level of within-category impedance. There is the strongest cross-vehicle effect for SC density, with $\beta_1$ = -13.041, which proves that cars are a major cause of speed disruption for utility vehicles. Three-wheelers also have a significant impact, with $\beta_4$ = -8.232, whereas NMVs have the smallest effect, $\beta_6$ = -0.487. The residual variance of 0.724 is the lowest in the system, meaning that utility vehicle speeds are modelled with the highest accuracy and show fairly consistent behaviour.

**Heavy Vehicles ($V_{HV}$).** The intercept of 10.276 m/s, or about 37.0 km/h, means that the theoretical free-flow speed of heavy vehicles lies between that of small cars and that of utility vehicles. The self-density coefficient, $\beta_3$ = -7.159, together with a narrow range of credible values between -7.540 and -6.777, shows that there is a strong and accurately estimated within-category impedance. Since heavy vehicles take up a lot of road space and are unable to move around one another easily, this is significant. The density of small cars has the main effect between vehicles, with $\beta_1$ equal to -35.157, showing that the high volume and lateral movement of cars greatly disrupt the operation of heavy vehicles. Three-wheelers and two-wheelers also reduce the speed of heavy vehicles, with coefficients of -12.391 and -6.196 respectively. This is in agreement with smaller vehicles moving into the gaps in front of or alongside heavy vehicles. The residual variance, $\sigma^2$ = 2.087, shows a moderate amount of unexplained variation.

**Three-Wheelers ($V_{3W}$).** The speed of 10.351 m/s, which is about 37.3 km/h, is comparable to that of heavy vehicles, even though the fundamental behaviour is different. Three-wheelers can reach this base speed by constantly searching for gaps rather than by steadily staying in their lane. Their self-density coefficient, $\beta_4$ = -22.819, is high and shows that there is serious mutual interference as the density of three-wheelers increases. By gathering up near intersections and by the kerbs, they may cause repeated blocking and thus greatly reduce their own speeds. SC density continues to be the most significant factor across categories, with a self-density coefficient of $\beta_1$ = -32.638 and a narrow range of credible values from [-34.139, -31.138]. The effect of two-wheelers, $\beta_5$ = -7.665, is also considerable, implying that motorcycles and three-wheelers compete for similar lateral gaps. The residual variance of 1.462 shows that the equation is comparatively well fitted.

**Two-Wheelers ($V_{2W}$).** The intercept of 9.899 m/s, or about 35.6 km/h, shows the capacity of motorcycles to make use of small gaps while still being affected by the conditions that prevail in urban environments. The self-density coefficient, $\beta_5$ = -12.708, with a credible interval ranging from -13.752 to -11.663, demonstrates strong mutual speed suppression since motorcycles tend to concentrate in narrow lateral spaces. Once again, SC density is responsible for the main cross-vehicle effect, with $\beta_1$ equal to -28.033. The three-wheeler coefficient, $\beta_4$ = -11.732, is also substantial, indicating a strong level of competition between these two highly manoeuvrable types of vehicle for the same gaps. NMV density has a smaller but accurately estimated effect, $\beta_6$ = -1.670, the credible interval being [-1.868, -1.472]. The residual variance of 1.246 is the second lowest in the system and this means that the predictions are highly consistent.

**Non-Motorized Vehicles ($V_{NMV}$).** The NMV equation exhibits the highest level of behavioural and statistical complexity. Its intercept of 10.578 m/s is greater than the speeds typically observed for NMV and must therefore be interpreted with caution. It could be due to the extrapolation of the simultaneous system to densities approaching zero rather than representing a directly observed operating condition. SC density has the biggest impact across vehicles, with $\beta_1$ equal to -42.091 and a credible interval of [-44.256, -39.926], which confirms that cars are the main cause of the reduction in NMV speed. The effect of two-wheelers is also significant, with $\beta_5$ being -9.138. The coefficient for HV is exceptional: $\beta_3$ is -0.001, with a credible interval of [-0.462, 0.459] and P(positive) equal to 0.498. The posterior distribution is essentially centred on zero, showing that there is no reliable directional effect of HV density on NMV speed. This is in contrast to the MLE estimate of +0.386, which had suggested that NMVs might speed up when near heavy vehicles. The Bayesian result does not support that interpretation since the relationship is too uncertain and variable. Furthermore, the NMV equation has the largest residual variance, $\sigma^2 = 3.044$, which indicates the presence of different types of NMV, poor lane discipline, and highly variable interactions with motorised traffic.

## RESULTS AND DISCUSSION

### Effect of Traffic Volume on Different Traffic Composition Scenarios

The developed simultaneous speed prediction models were used to evaluate the influence of traffic volume on the operating speeds of individual vehicle categories under five representative heterogeneous traffic compositions (**Figures 1–5**). The simultaneous equation system was solved for traffic volumes ranging from 1,000 to 7,000 veh/h at 500 veh/h intervals while varying the proportions of standard cars (SC), utility vehicles (UV), heavy vehicles (HV), three-wheelers (3W), two-wheelers (2W), and non-motorized vehicles (NMV).

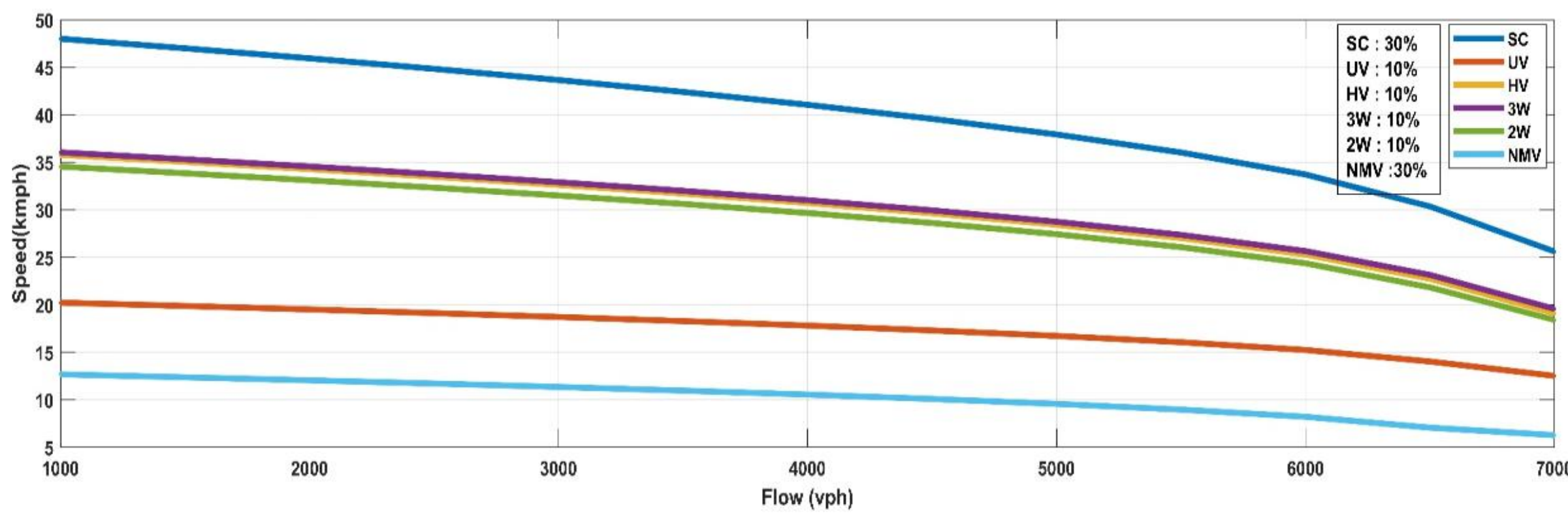

**Figure 1** Variation in Speed of Individual Vehicle with Traffic Volume (composition of traffic stream was kept fixed with SC, UV, HV, 3W, 2W, and NMV as 30%, 10%, 10%, 10%, 10%, and 30%, respectively)

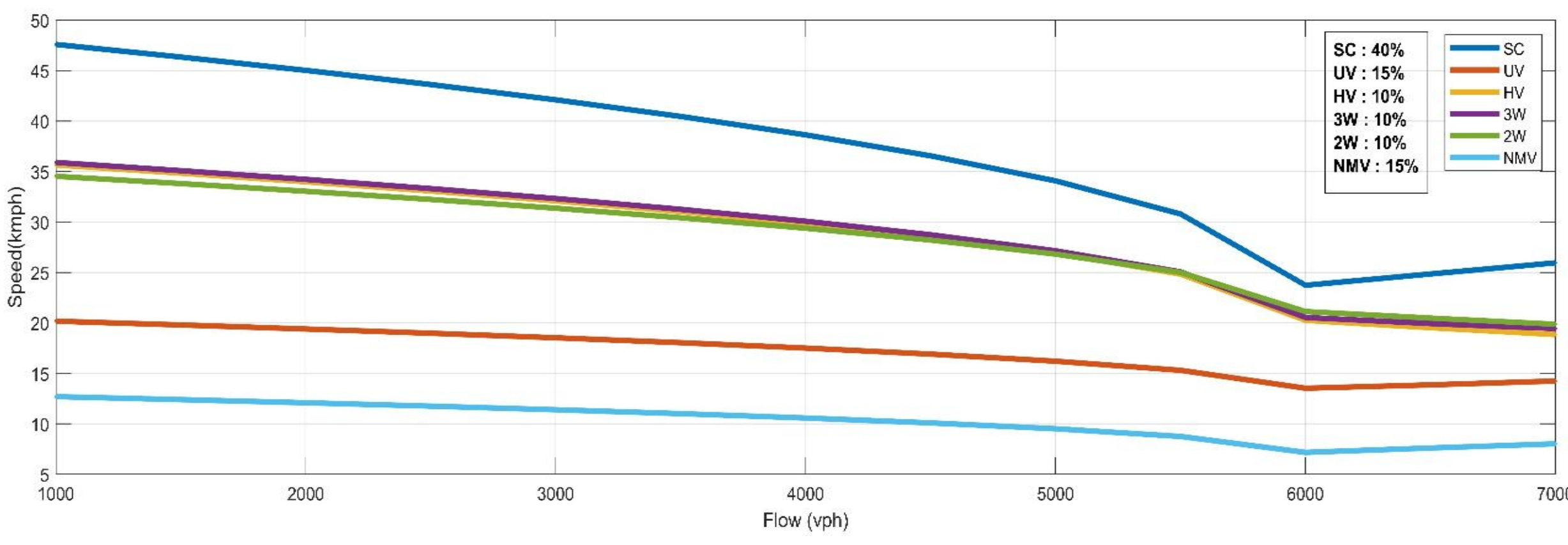

**Figure 2** Variation in Speed of Individual Vehicle with Traffic Volume (composition of traffic stream was kept fixed with SC, UV, HV, 3W, 2W, and NMV as 40%, 15%, 10%, 10%, 10%, and 15%, respectively)

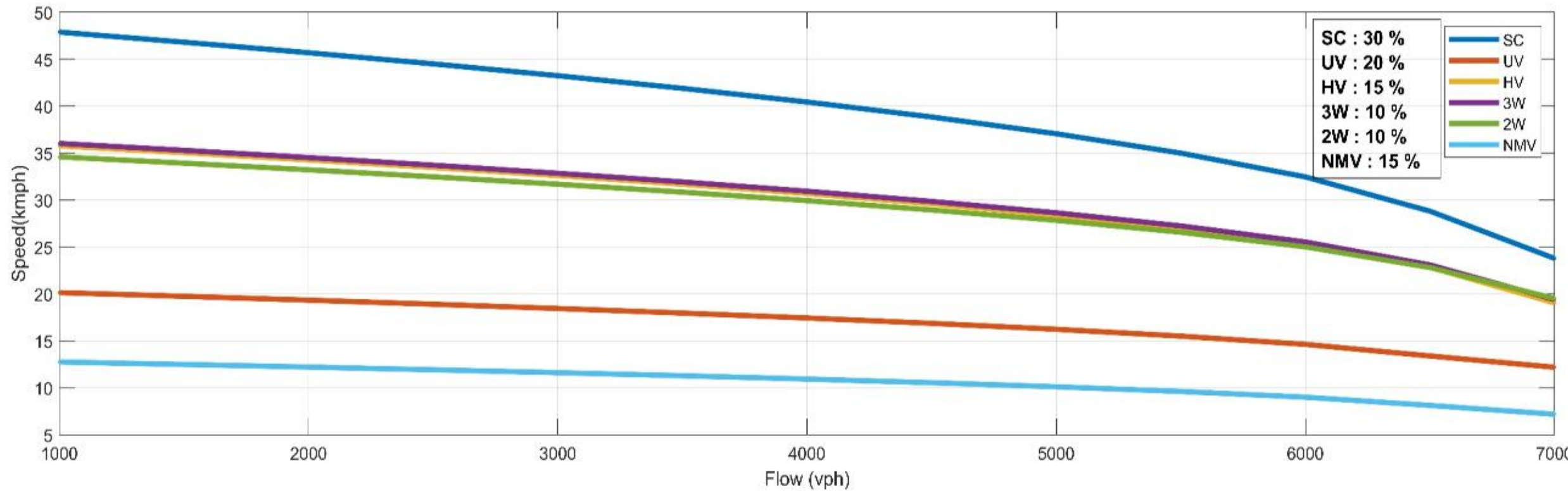

**Figure 3** Variation in Speed of Individual Vehicle with Traffic Volume (composition of traffic stream was kept fixed with SC, UV, HV, 3W, 2W, and NMV as 30%, 20%, 10%, 15%, 15%, and 10%, respectively)

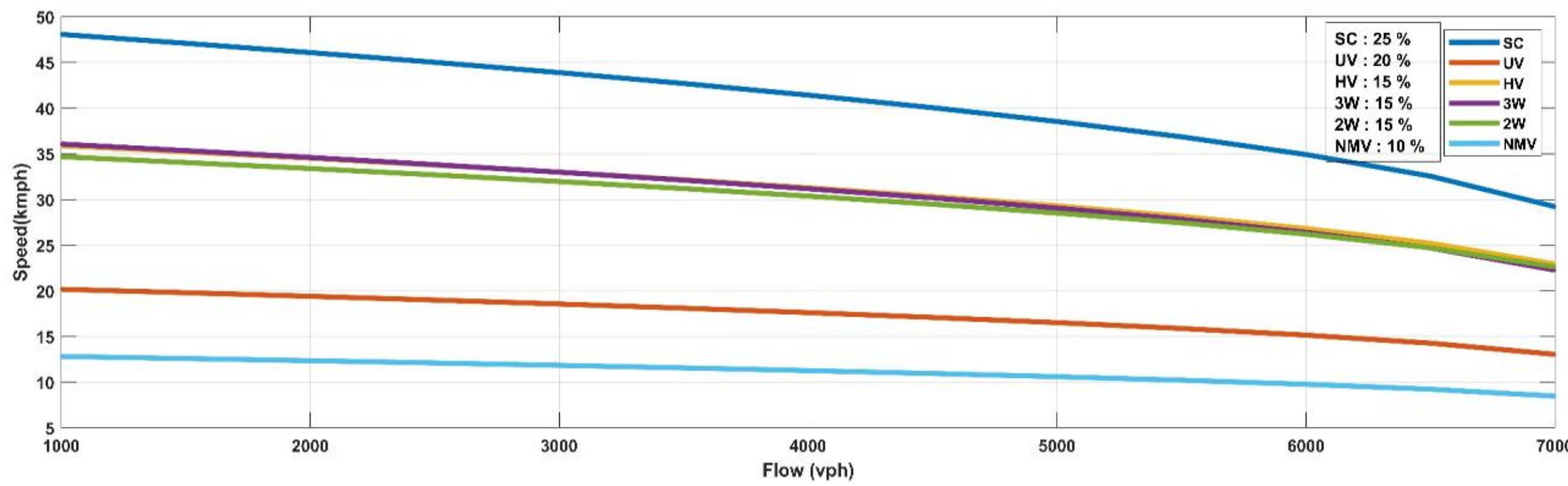

**Figure 4** Variation in Speed of Individual Vehicle with Traffic Volume (composition of traffic stream was kept fixed with SC, UV, HV, 3W, 2W, and NMV as 25%, 20%, 15%, 15%, 15%, and 10%, respectively)

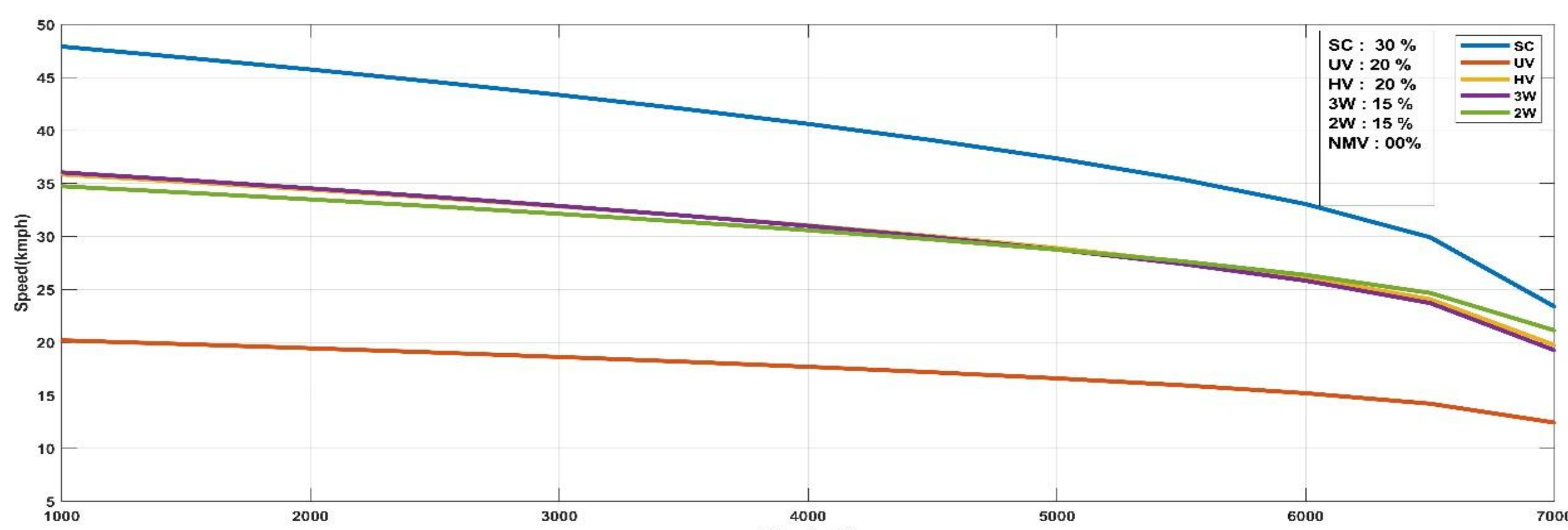

**Figure 5** Variation in Speed of Individual Vehicle with Traffic Volume (composition of traffic stream was kept fixed with SC, UV, HV, 3W, 2W, and NMV as 30%, 20%, 20%, 15%, 15%, and 00% respectively)

In all the different traffic composition scenarios, the operating speed fell continuously as the traffic volume increased for each type of vehicle. This is in keeping with the basic speed–flow relationship, since greater demand leads to more frequent vehicle interactions, fewer opportunities for maneuvering, and therefore lower operating speeds. The fact that the proposed simultaneous model takes into account the interactions between vehicle categories means that the speed variations it predicts include both the level of traffic demand and the heterogeneous nature of the traffic composition.

Standard cars were able to keep the highest operating speeds since they had better maneuverability and acceleration, while non-motorized vehicles had the lowest speeds and were most sensitive to rising traffic demand. Utility vehicles and heavy vehicles also had relatively low operating speeds as a result of their bigger size and limited maneuverability. On the other hand, two-wheelers and three-wheelers generally occupied a middle position, gaining from improved lateral maneuverability but still being affected by increasing congestion.

Even though the general relationship between speed and flow stayed the same, the make-up of the traffic had a significant effect on the amount of speed reduction. With the baseline composition (30% SC, 10% UV, 10% HV, 10% 3W, 10% 2W, and 30% NMV; **Figure 1**), every type of vehicle showed a steady decrease in speed as traffic demand increased. When the percentage of SC was raised to 40% and the proportion of NMV was reduced (**Figure 2**), speeds were comparatively higher at low and moderate traffic volumes; but beyond about 5,500 vehicles per hour speeds dropped sharply, showing that severe congestion had set in.

As can be seen in **Figures 3** and **4**, the greater the proportion of UV, HV, 3W, and 2W vehicles, the more significant the slowdowns become, especially when traffic volume is high. This is because these types of vehicles either take up more space on the road or travel at lower speeds, which in turn leads to more interactions between vehicles and reduces the effective capacity available to other vehicles.

Because of the lack of NMVs (as shown in **Figure 5**), there were small gains in operating speeds for all categories of motorised vehicles and the timing of severe congestion was postponed. By reducing the number of slow-moving vehicles, the differences in speed within the traffic flow were lessened and traffic stability was improved. However, operating speeds still fell as traffic volume increased, which showed that traffic demand had remained the main factor determining traffic performance regardless of the mix of vehicle types.

**Effect of Traffic Composition on Vehicle Speeds**

In order to separate out the effect of the make-up of traffic from that of traffic demand, the simultaneous speed prediction model was solved with a constant traffic volume of 4,500 vehicles per hour while gradually changing the proportions of the different types of vehicles. Further analyses varied both the traffic volume and the composition of the vehicles at the same time in order to look at their combined impact on operating speeds (**Figures 6–9**).

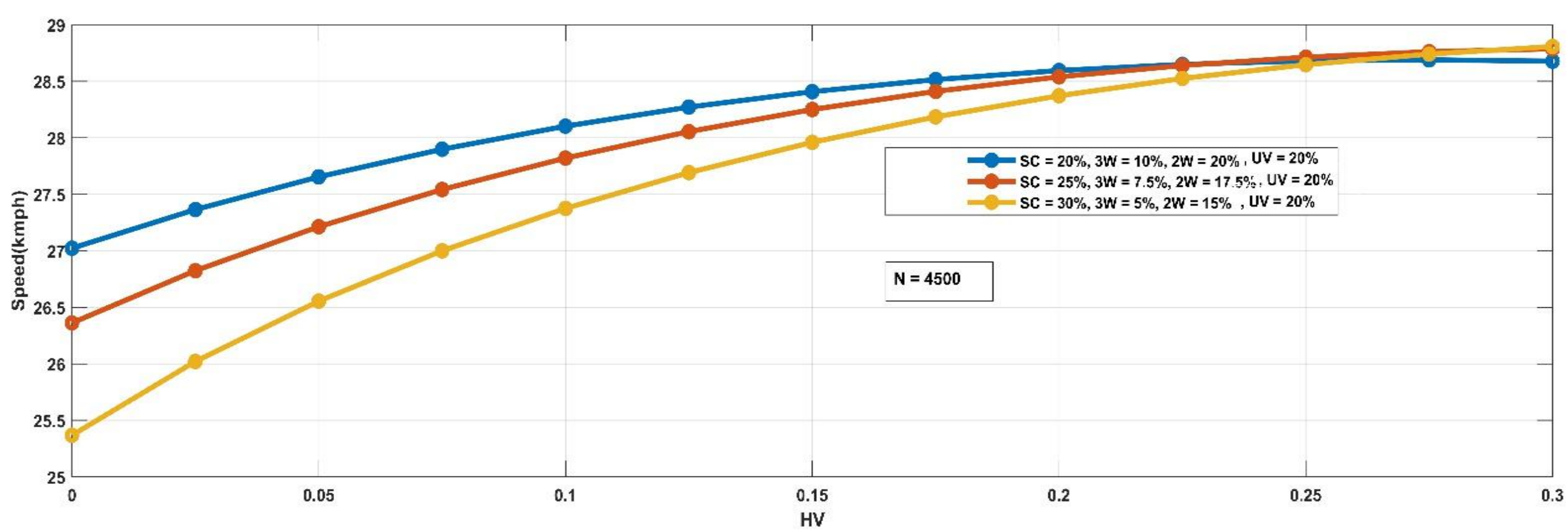

**Figure 6** Variation in Speed of Heavy Vehicle with Traffic Composition

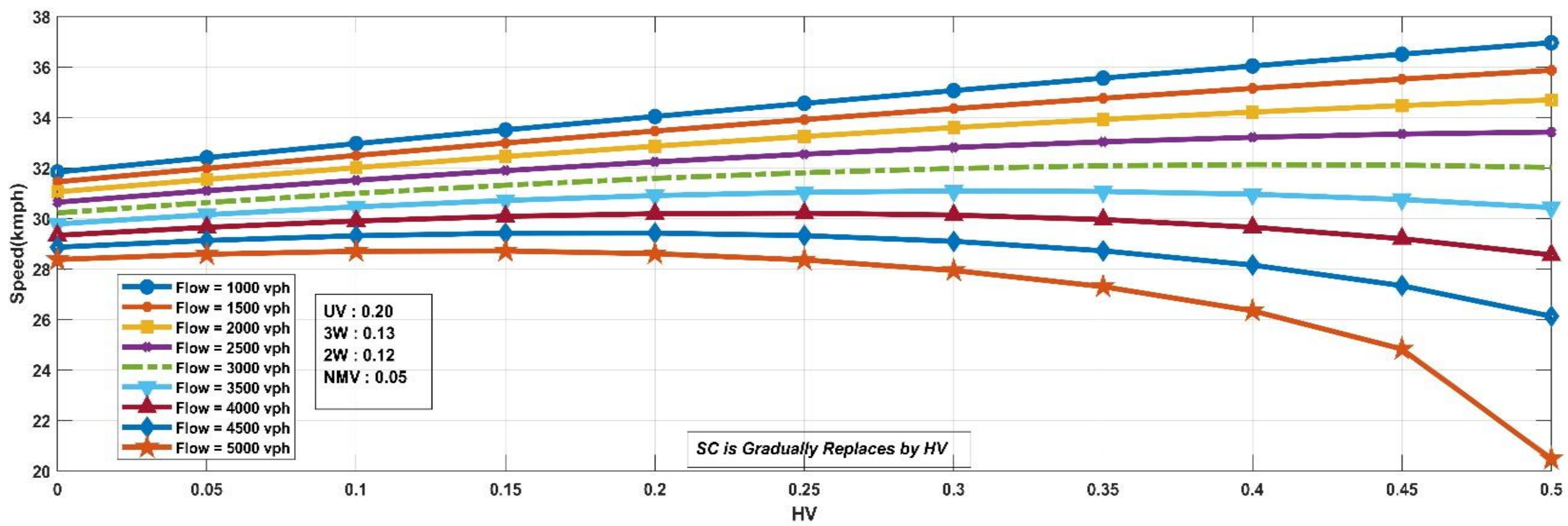

**Figure 7** Speed of Heavy Vehicle with volume and composition where Standard Car is gradually replaced by Heavy Vehicle.

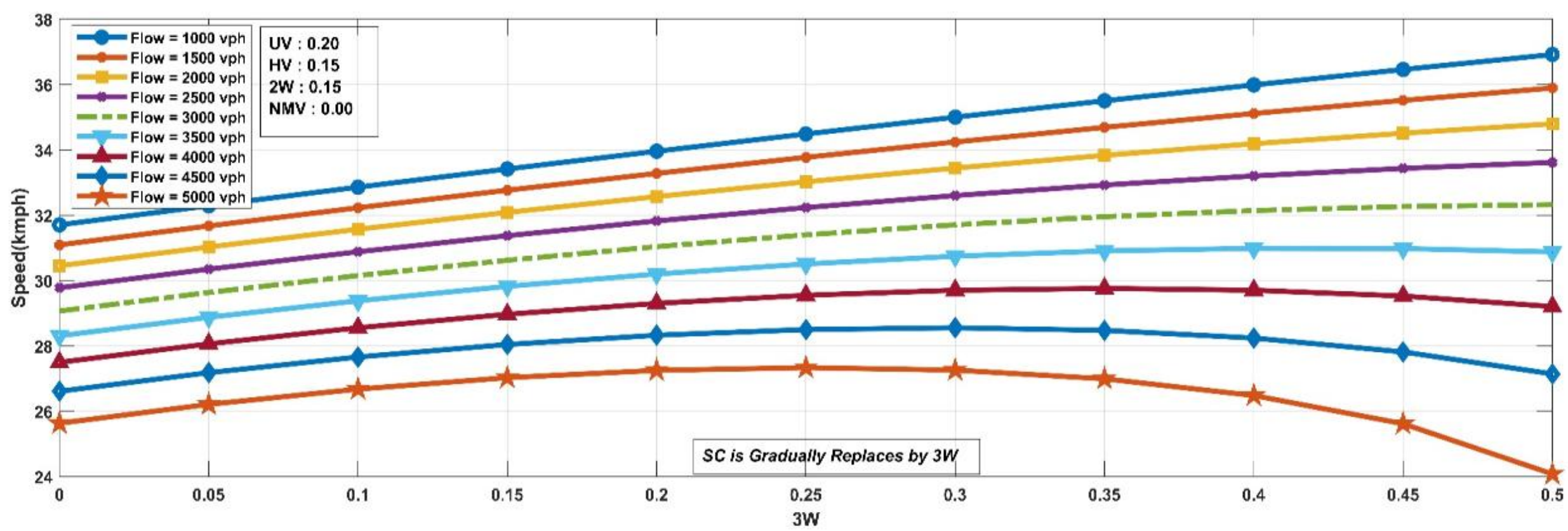

**Figure 8** Speed of Three Wheeler with volume and composition, where Three Wheeler gradually replaces Standard Car.

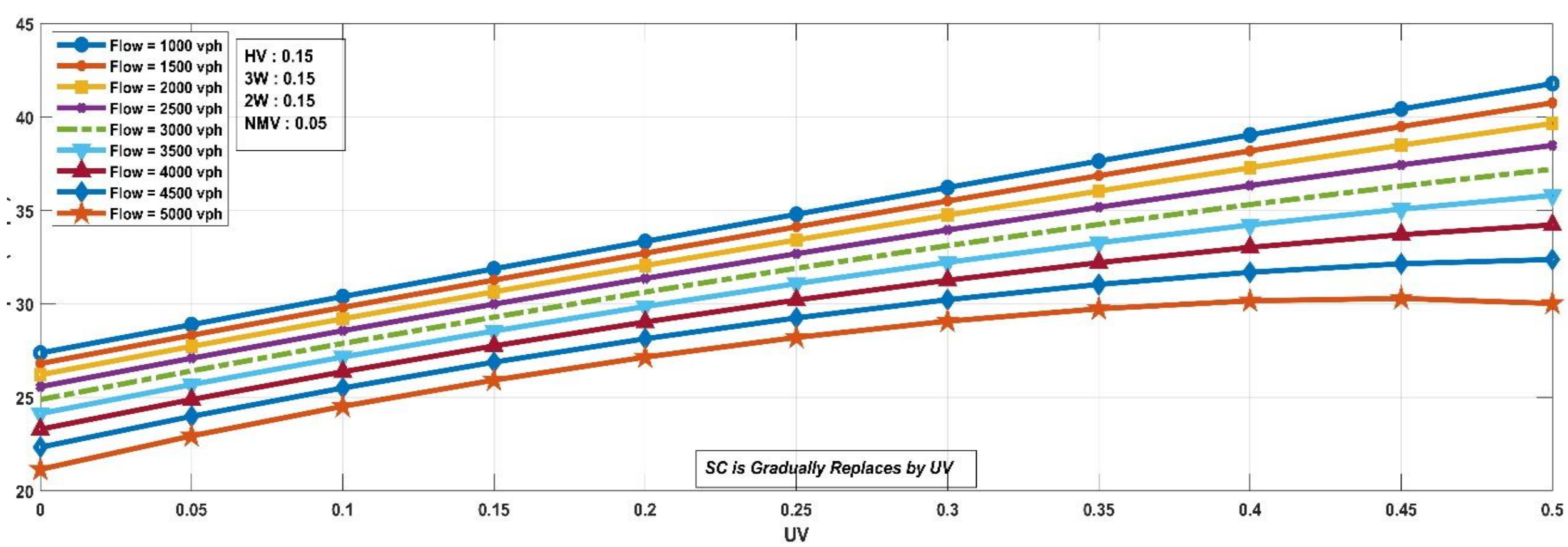

**Figure 9** Speed of Utility Vehicle with volume and composition where Standard Car is gradually replaced by Utility Vehicle.

**Figure 6** shows the effect of the proportion of heavy vehicles under three typical traffic composition scenarios. Although heavy vehicles are larger in size and less manoeuvrable, their predicted operating speed increases slightly as the proportion of heavy vehicles rises. This occurs because of the simultaneous nature of the model: when the share of heavy vehicles increases, the proportion of smaller and more laterally active vehicles decreases, leading to fewer instances of overtaking and weaving. Nevertheless, the actual speed levels vary among the three scenarios, which shows that the overall traffic composition, and not just the proportion of one type of vehicle, determines traffic performance.

**Figures 7–9** demonstrate the combined influence of traffic volume and traffic composition. Under low traffic volumes, changes in vehicle composition have only a modest effect because sufficient roadway space allows relatively unrestricted vehicle movements. As traffic demand increases, however, composition becomes increasingly influential as competition for available roadway space intensifies and interactions among vehicle categories become more frequent.

The effect of gradually substituting standard cars with heavy vehicles while keeping the proportions of the other vehicle types unchanged is shown in **Figure 7**. When traffic volumes are low or moderate, a moderate rise in the share of heavy vehicles results in an improvement of operating speeds since it reduces the interactions involving standard cars; but in congested conditions higher proportions of heavy vehicles cause considerable decreases in speed due to their greater take up of space, lower ability to accelerate, and limited maneuverability.

A similar nonlinear relationship is observed when standard cars are gradually replaced by three-wheelers (**Figure 8**). Moderate three-wheeler proportions improve traffic performance, particularly under uncongested conditions, owing to their compact size and greater lateral maneuverability. Under high traffic volumes, however, excessive three-wheeler presence increases lateral interference and competition for roadway space, resulting in reduced operating speeds.

On the other hand, increasing the share of utility vehicles (as shown in **Figure 9**) results in a steady improvement in operating speeds over the range of traffic volumes studied. Utility vehicles seem to produce relatively stable interactions within the mixed traffic flow, which in turn leads to more smooth traffic movement. While the amount of improvement diminishes under congested conditions, the positive trend continues to hold. These results show that the makeup of traffic has a significant effect on operating speeds, especially as traffic demand nears the capacity of the road, and thus it is important to take into account vehicle heterogeneity in macroscopic speed prediction models.

**Transferability Assessment of the Proposed Model**

The ability of the Bayesian simultaneous speed prediction model to be applied elsewhere was tested by using two separate urban primary road sections—Sayed Mahbub Morshed Road and Kazi Nazrul Islam Avenue—which had not been employed during the development of the model. Although these road sections have geometric features similar to those of the calibration sites, they differ in terms of traffic makeup and operating conditions, thus allowing for an independent evaluation of the model's generalizability. The UAV videos were processed with the same DEEGITS framework in order to obtain the hourly traffic volumes, the compositions of the vehicles, and the space-mean speeds for the six vehicle categories.

The predicted speeds were compared with the speeds observed in the field, and the model's performance was evaluated by means of the distribution of prediction errors and various statistical measures of reliability. As **Figure 10** illustrates, the Bayesian model in question always yielded narrower error distributions and smaller median prediction errors for all categories of vehicle, showing that it had improved predictive accuracy and robustness when faced with traffic conditions it had not seen before. This enhancement was especially clear in the case of standard cars, heavy vehicles, and non-motorized vehicles, since these types of vehicles generally exhibit greater variability due to the complex interactions that take place in heterogeneous traffic. The Bayesian approach was able to capture these uncertainties by estimating the posterior parameter distributions, which in turn led to more stable speed predictions.

The prediction errors were least for utility vehicles, three-wheelers, and two-wheelers since their operating characteristics did not change much from one of the validation sites to another. In contrast, standard cars and heavy vehicles showed a considerably greater degree of prediction variability, this being due to their greater sensitivity to changes in traffic composition and congestion. Even though non-motorized vehicles had a high level of variability during the model development phase, the Bayesian framework greatly reduced the prediction uncertainty, thus showing that it is capable of representing heterogeneous, non-lane-based traffic.

The statistical reliability of the model was investigated further by looking at how the t-statistic and its associated p-value varied over the range of observed speeds (see **Figure 11**). When speeds were very low, the higher p-values and lower t-statistics show that there is greater uncertainty caused by stop-and-go traffic and by frequent interactions between vehicles. As speed increased, prediction reliability improved quickly since the p-values dropped below the usual significance level and the t-statistics rose. The fact that p-values remained consistently low at moderate and higher operating speeds proves that the proposed Bayesian simultaneous speed prediction model is statistically significant and that its predictive performance is stable under a variety of traffic conditions.

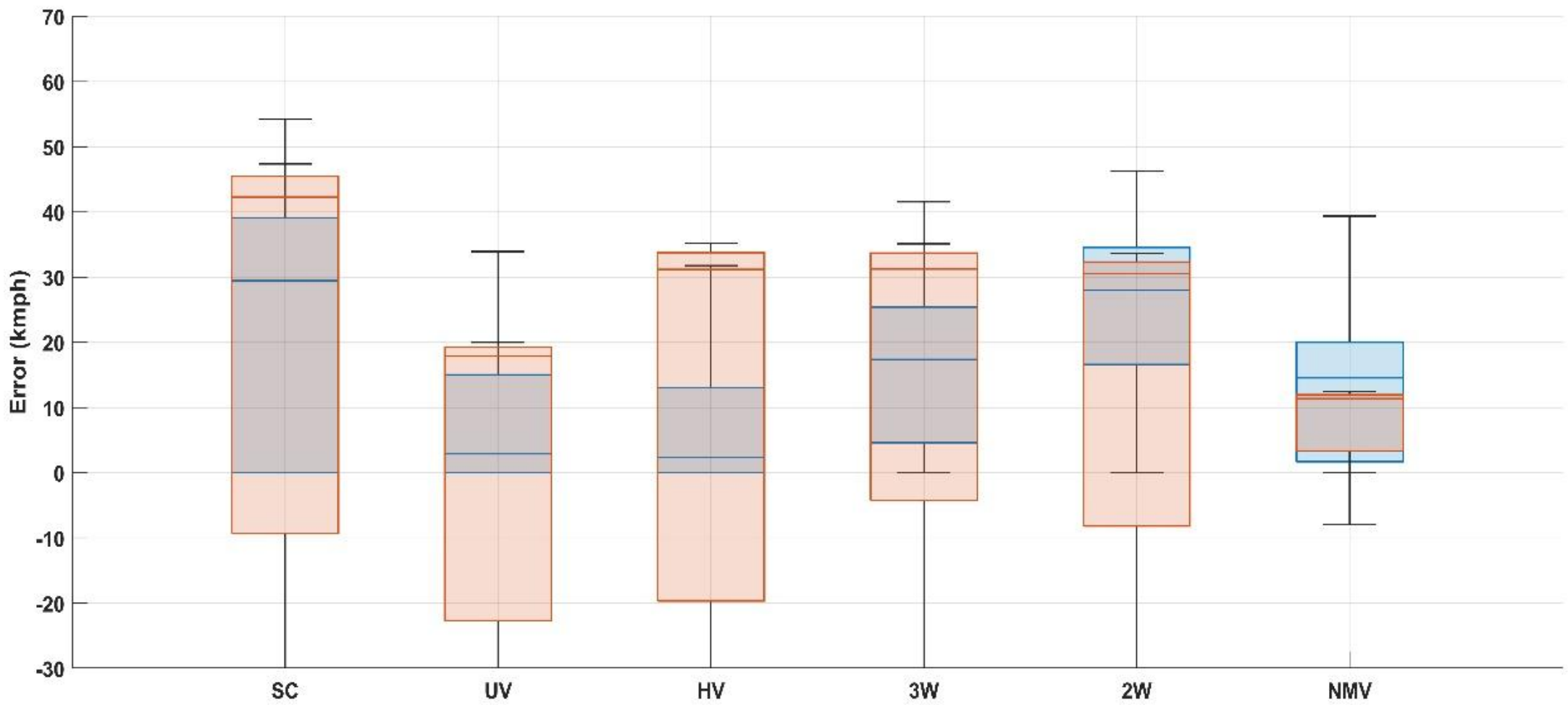

**Figure 10 Error between observed and predicted speed**

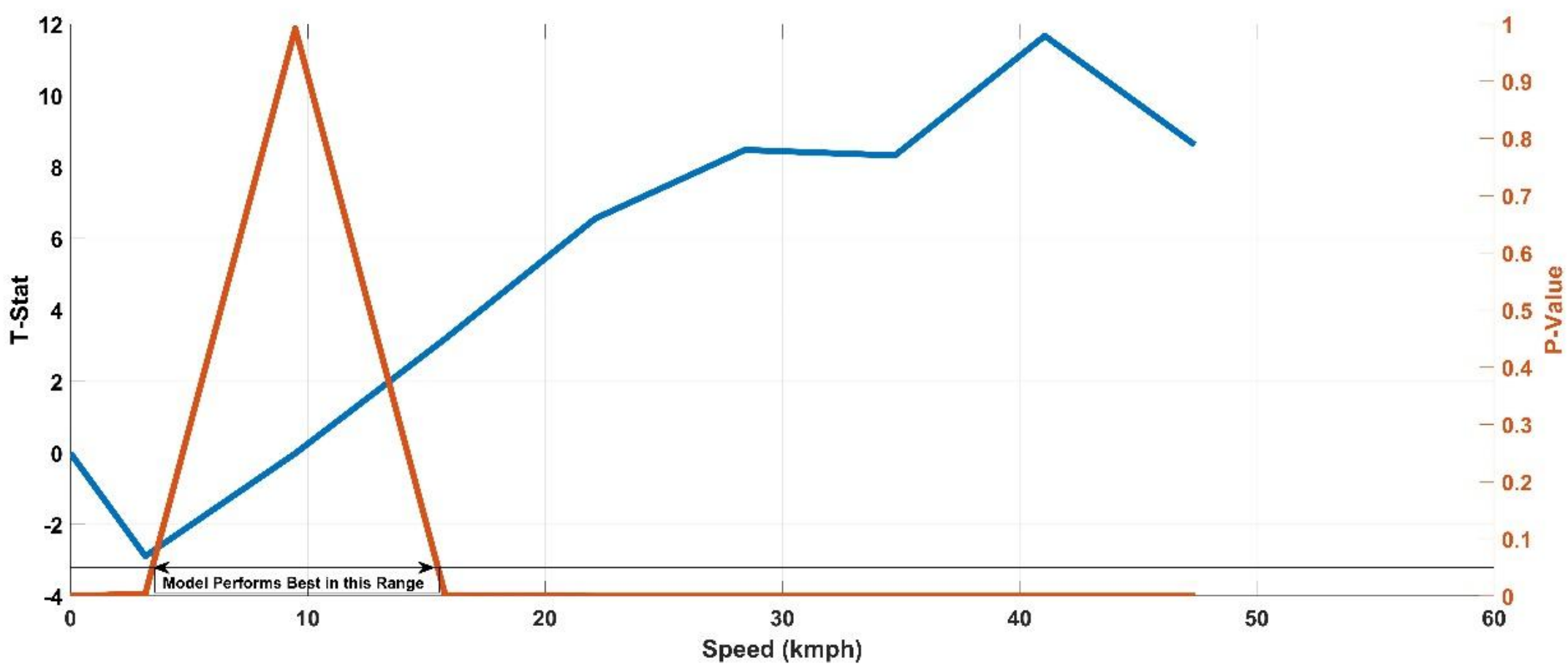

**Figure 11 T-State and P-Value of the Transferability of Simultaneous Equation Optimization**

**Practical Applications**

The proposed Bayesian simultaneous speed prediction framework provides a practical decision-support tool for evaluating traffic performance on heterogeneous, non-lane-based urban roads. By explicitly accounting for interactions among multiple vehicle categories, the model enables transportation agencies to assess the operational impacts of changes in traffic demand and vehicle composition. The framework can support traffic management, roadway performance evaluation, traffic simulation, and infrastructure planning by providing reliable speed estimates under varying operating conditions. Because the methodology is transferable to similar heterogeneous traffic environments, it offers a useful analytical tool for urban road networks in low- and middle-income countries where mixed traffic and weak lane discipline prevail.

**CONCLUSION AND RECOMMENDATIONS**

A Bayesian simultaneous speed prediction method was developed in this study for use in heterogeneous, non-lane-based urban traffic, the approach making use of traffic data obtained from UAVs along five major road corridors in Dhaka. Different from traditional speed–flow models which assume homogeneous traffic conditions, the present framework explicitly shows the interactions between six types of vehicles—Standard Cars (SC), Utility Vehicles (UV), Heavy Vehicles (HV), Three-Wheelers (3W), Two-Wheelers (2W), and Non-Motorized Vehicles (NMV)—by means of a set of simultaneous speed equations estimated with the Markov Chain Monte Carlo (MCMC) method. The Bayesian approach made it possible to obtain the full posterior distributions of the model parameters, thus allowing for reliable parameter estimation as well as for the assessment of uncertainty.

The models developed showed a high ability to make predictions for all types of vehicles. The coefficient of determination ($R^2$) was in the range 0.922 to 0.956, and the corresponding RMSE values were between 0.50 and 1.56 m/s. Of the six models, the one for two-wheelers achieved the greatest predictive accuracy ($R^2 = 0.956$), while the model for heavy vehicles had the worst goodness-of-fit yet it was still acceptable ($R^2 = 0.922$). This shows that the modeling framework presented is able to capture the complex speed–density relationships that govern heterogeneous traffic streams.

When the six coupled equations were solved together, it was found that an increase in traffic volume always led to a decrease in the operating speeds of all types of vehicles, the extent of this decrease however depending strongly on the make-up of the vehicles. Standard cars kept the highest operating speeds whatever the traffic conditions under investigation, while non-motorized vehicles suffered the largest reductions in speed as traffic demand rose. Changes in traffic composition also showed that the composition of the vehicle mix has a significant effect on the performance of the traffic stream. A greater share of maneuverable vehicles produced a different kind of interaction pattern compared to that of larger vehicles, and when non-motorized vehicles were removed the operating speeds increased and congestion was delayed in its onset, showing how much slow-moving traffic affects mixed-flow operations.

An analysis of transferability using separate sections of road showed that the Bayesian simultaneous model in question generalised very well beyond the calibration dataset. For all vehicle types, the prediction errors were consistently lower than those achieved with the benchmark model, and statistical evaluation proved that the model had a significant predictive ability over the entire range of practical operating speeds. This shows that the proposed approach has enough robustness to be applied to urban primary roads which have similar heterogeneous traffic characteristics.

The modeling framework that has been proposed serves as a practical means of estimating operating speeds specific to different categories when traffic demand and composition vary, and it can be readily applied to traffic simulation, operational analysis, roadway capacity evaluation, and intelligent traffic management systems in developing countries where heterogeneous and poorly lane-disciplined traffic is the norm. Further research could extend the framework by taking into account roadway geometric features, environmental factors, driver behavioural characteristics, and the interactions between connected and autonomous vehicles in order to enhance its usefulness in changing mixed-traffic situations.

## AUTHOR CONTRIBUTIONS

The authors confirm contribution to the paper as follows: study conception and design: S.M. T. Alam, N. Haque, M. A. Raihan, M. Hadiuzzaman; data collection: S.M. T. Alam; analysis and interpretation of results: S.M. T. Alam, N. Haque, M. A. Raihan, M. M. Shahrier; draft manuscript preparation: M. M. Shahrier, N. Haque, M.A. Raihan. All authors reviewed the results and approved the final version of the manuscript.

## DECLARATION OF CONFLICTING INTERESTS

The authors declared no potential conflicts of interest with respect to the research, authorship, and/or publication of this article.

## FUNDING

This research work is supported by the Committee for Advanced Studies and Research (CASR) of the Bangladesh University of Engineering and Technology (BUET).

## REFERENCES

Ahmed, Afzal, Dong Ngoduy, Muhammad Adnan, and Mirza Asad Ullah Baig. 2021. "On the Fundamental Diagram and Driving Behavior Modeling of Heterogeneous Traffic Flow Using UAV-Based Data."

*Transportation Research Part A: Policy and Practice* 148: 100–115. https://doi.org/10.1016/j.tra.2021.03.001.

Alomari, Ahmad H., Taisir S. Khedaywi, Ahmad R. O. Marian, and Ali A. Jadah. 2022. "Traffic Speed Prediction Techniques in Urban Environments." *Heliyon* 8 (12).

Brooks, Stephen. 1998. "Markov Chain Monte Carlo Method and Its Application." *Journal of the Royal Statistical Society: Series D (The Statistician)* 47 (1): 69–100.

Dey, Partha Pratim, Satish Chandra, and S. Gangopadhaya. 2006. "Speed Distribution Curves under Mixed Traffic Conditions." *Journal of Transportation Engineering* 132 (6): 475–481.

Dhamaniya, Ashish, and Satish Chandra. 2013. "Speed Prediction Models for Urban Arterials under Mixed Traffic Conditions." *Procedia - Social and Behavioral Sciences* 104: 342–351. https://doi.org/10.1016/j.sbspro.2013.11.127.

Donnell, Eric T., Yingwei Ni, Michelle Adolini, and Lily Elefteriadou. 2001. "Speed Prediction Models for Trucks on Two-Lane Rural Highways." *Transportation Research Record* 1751 (1): 44–55. https://doi.org/10.3141/1751-06.

Du, Yiheng, Xiao Qin, Zhaobin Jia, Kai Yu, and Ming Lin. 2021. "Traffic Speed Prediction Based on Heterogeneous Graph Attention Residual Time Series Convolutional Networks." *AI* 2 (4): 650–661. https://doi.org/10.3390/ai2040039

Greenshields, Bruce D., J. Rowland Bibbins, W. S. Channing, and Harvey H. Miller. 1935. "A Study of Traffic Capacity." *Highway Research Board Proceedings* 14 (1): 448–477.

Haque, Nazmul, Md Asif Raihan, Md Mizanur Rahman, and Md Hadiuzzaman. 2024. "A Review on Following Behavioral Models: Regular to Connected Autonomous Vehicle Heterogeneity." *IATSS Research* 48 (2): 163–179. https://doi.org/10.1016/j.iatssr.2024.03.002.

Himes, Scott C., and Eric T. Donnell. 2010. "Speed Prediction Models for Multilane Highways: Simultaneous Equations Approach." *Journal of Transportation Engineering* 136 (10): 855–862. https://doi.org/10.1061/(ASCE)TE.1943-5436.0000149.

Islam, Muttahirul, Nazmul Haque, and Md. Hadiuzzaman. 2024. "DEEGITS: Deep Learning Based Framework for Measuring Heterogeneous Traffic State in Challenging Traffic Scenarios." Paper presented at the 103rd Annual Meeting of the Transportation Research Board, Washington, DC. https://doi.org/10.48550/arXiv.2411.08335.

Khan, Muhammad Arsalan, Wim Ectors, Tom Bellemans, Davy Janssens, and Geert Wets. 2017. "UAV-Based Traffic Analysis: A Universal Guiding Framework Based on Literature Survey." *Transportation Research Procedia* 22: 541–550. https://doi.org/10.1016/j.trpro.2017.03.043.

Liao, Binbing, Jingqing Zhang, Chao Wu, Douglas McIlwraith, Tong Chen, Shengwen Yang, Yike Guo, and Fei Wu. 2018. "Deep Sequence Learning with Auxiliary Information for Traffic Prediction." In *Proceedings of the 24th ACM SIGKDD International Conference on Knowledge Discovery & Data Mining*.

Maini, P., and S. V. Khan. 2000. "Discharge Characteristics of Heterogeneous Traffic at Signalized Intersections." *Transportation Research Circular*.

Monteil, Julien, Andrei Dekusar, Claudio Gambella, Yassine Lassoued, and Martin Mevissen. 2021. "On Model Selection for Scalable Time Series Forecasting in Transport Networks." *IEEE Transactions on Intelligent Transportation Systems* 23 (7): 6699–6708.

Munawar, Ahmad. 2011. "Speed and Capacity for Urban Roads: Indonesian Experience." *Procedia - Social and Behavioral Sciences* 16: 382–387. https://doi.org/10.1016/j.sbspro.2011.04.459.

Najjar, Yacoub M., Robert W. Stokes, and E. R. Russell. 2000. "Setting Speed Limits on Kansas Two-Lane Highways: Neuronet Approach." *Transportation Research Record* 1708 (1): 20–27. https://doi.org/10.3141/1708-03.

Sarna, A. C., P. K. Jain, and G. Chandra. 1989. "Capacity of Urban Roads—A Case Study of Delhi and Bombay." *Highway Research Bulletin* (New Delhi) 40.

Tan, Dang Minh, Nguyen Hoang Tung, and Bui Xuan Cay. 2021. "Maximum Likelihood Estimation Method for Speed Prediction of Vehicles in Mixed Traffic Condition." In *Proceedings of the 3rd International Conference on Sustainability in Civil Engineering (ICSCE 2020)*, 359–365. Singapore. https://doi.org/10.1007/978-981-16-0053-1_45.

Turner, Shawn M., William L. Eisele, Robert J. Benz, and Douglas J. Holdener. 1998. *Travel Time Data Collection Handbook*. Washington, DC: Federal Highway Administration, U.S. Department of Transportation.

Wang, Chien-Yao, Alexey Bochkovskiy, and Hong-Yuan Mark Liao. 2022. "YOLOv7: Trainable Bag-of-Freebies Sets New State-of-the-Art for Real-Time Object Detectors." arXiv preprint. https://doi.org/10.48550/arXiv.2207.02696.

Williams, JAMES C. 1997. "Macroscopic Flow Models." *Revised Monograph on Traffic Flow Theory*, 1–31.

Yoon, Seonkyoo, Woncheol Cho, Jun-Haeng Heo, and Chul Eung Kim. 2010. "A Full Bayesian Approach to Generalized Maximum Likelihood Estimation of Generalized Extreme Value Distribution." *Stochastic Environmental Research and Risk Assessment* 24 (5): 761–770. https://doi.org/10.1007/s00477-009-0362-7.

Yusuf Ali, Md. 2021. *Stochastic Modeling of Equilibrium Speed-Density Relationship for Non-Lane-Based Heterogeneous Traffic*. Master's thesis, Bangladesh University of Engineering and Technology, Dhaka, Bangladesh. http://lib.buet.ac.bd:8080/xmlui/handle/123456789/5930.